\documentclass[aps,twocolumn,showlabels,showrefs,amsmath,amssymb,pre,superscriptaddress, floatfix, colors, longbibliography]{revtex4-1}

\usepackage{graphicx}
\usepackage{amssymb}
\usepackage{color}
\usepackage{amsmath}
\usepackage{textcomp}
\usepackage{amsfonts}
\usepackage{hyperref}

\definecolor{mycolor}{RGB}{50,0,255}

\newcommand{\av}[1]{\langle #1 \rangle}

\newcommand{\FigPath}{./}

\begin{document}
\title{Flocking and spreading dynamics in populations of self-propelled agents}


\author{Demian Levis}
\thanks{These two authors contributed equally}
\affiliation{CECAM  Centre  Europ\'eeen  de  Calcul  Atomique  et  Mol\'eculaire,
Ecole  Polytechnique  F\'ed\'erale  de  Lausanne,  Batochime,  Avenue  Forel  2,  1015  Lausanne,  Switzerland}
\affiliation{Departament de Fisica de la Materia Condensada, Universitat de Barcelona, Marti i Franques 1, 08028 Barcelona, Spain}
\affiliation{Universitat de Barcelona Institute of Complex Systems (UBICS), Universitat de Barcelona, 08028 Barcelona, Spain}
\author{Albert Diaz-Guilera}
\affiliation{Departament de Fisica de la Materia Condensada, Universitat de Barcelona, Marti i Franques 1, 08028 Barcelona, Spain}
\affiliation{Universitat de Barcelona Institute of Complex Systems (UBICS), Universitat de Barcelona, 08028 Barcelona, Spain}
\author{Ignacio Pagonabarraga}
\affiliation{CECAM  Centre  Europ\'eeen  de  Calcul  Atomique  et  Mol\'eculaire,
Ecole  Polytechnique  F\'ed\'erale  de  Lausanne,  Batochime,  Avenue  Forel  2,  1015  Lausanne,  Switzerland}
\affiliation{Departament de Fisica de la Materia Condensada, Universitat de Barcelona, Marti i Franques 1, 08028 Barcelona, Spain}
\affiliation{Universitat de Barcelona Institute of Complex Systems (UBICS), Universitat de Barcelona, 08028 Barcelona, Spain}
\author{Michele Starnini}
\thanks{These two authors contributed equally}
\email[corresponding author:]{michele.starnini@gmail.com}
\affiliation{ISI Foundation, via Chisola 5, 10126 Torino, Italy}


%
\begin{abstract}
Populations of self-propelled mobile agents -- animal groups, robot swarms or crowds of people -- that exchange information with their surrounding, host fascinating cooperative behaviors. While in many situations of interest the agents motion is driven by the transmission of information (e.g. the presence of an approaching predator) from neighboring peers, previous modeling efforts have mainly described situations where agents either sit on static networks, or move independently of the information spreading across the population. 
Here, we introduce a reference model to tackle this current lack of general framework. We consider mobile agents which align their direction of motion with their neighbors (based on the Kuramoto dynamics) 
 and carry an internal state governed by the Susceptible-Infected-Susceptible (SIS) epidemic process, characterizing the spread of information in the population, and affecting the way agents move in space. 
We show that the feedback between the agents motion and the information spreading process is responsible for (i) the enhancement of both flocking and information spreading, 
(ii) the emergence of complex spatial structures, or swarms, which can be controlled by the velocity of the agents. 
The SIS dynamics is able to drive a flocking phase transition even in the absence of explicit velocity-alignment interaction, recovering a behavior reminiscent of Vicsek-like systems but featuring macro-phase separation rather than micro-phase separation as the Vicsek model. 
We show that the formation of dense swarms at low velocities reduces the epidemic threshold of information spreading with respect to the mean field limit in which agents interact globally. By bridging together soft active matter physics and agent based modeling of complex systems, we shed light upon a general positive feedback mechanism that crucially affects the collective behavior of mobile agents, providing a reference framework to study realistic situations where this mechanism is at play. 
\end{abstract}

\pacs{-}
 

\date{\today}
\maketitle

\setlength{\textfloatsep}{10pt} 
\setlength{\intextsep}{10pt}



Mobile agents  -- animals, people, or robots -- may interact among them by exchanging information,
which directly influences the way they move through space. 
These interactions can trigger the emergence of fascinating collective states, such as the murmuration of starling flocks \cite{SumpterBook}, robot swarming \cite{Rubenstein2014}, or the collective motion of fish schools avoiding a predator's attack \cite{ioannou2012predatory}.
Individuals in a group may obtain information about the environment by observing the behavior of surrounding peers, to capture the presence of potential threats or opportunities.
For animals, information could be related to the location of a food source or an approaching predator. 
In shoaling fish, for instance, social cues transmitted through the shoal have been suggested to enhance early predator detection  \cite{Chicoli:2016aa}.  
In humans, the behavior of crowds in life-threatening situations can be determined by the spread of panic, transmitted locally among neighboring individuals \cite{Helbing:2000aa, PhysRevE.51.4282}. 

Populations of self-propelled agents, such as the ones mentioned above, constitute examples of active matter:  systems composed of entities which pump energy from their environment to perform motion  \cite{ramaswamy2010rev,MarchettiRev}. 
Interactions between self-propelled agents trigger the emergence of a plethora of complex non-equilibrium self-organized states with no counterpart in passive systems, from the collective motion of grains \cite{Deseigne2010, Weber2013} and coordinate migration of cancer cells \cite{deisboeck2009collective},  down to the dynamical clustering of natural (e.g. bacteria) or synthetic (e.g. active colloids) microswimmers \cite{BechingerRev}. 
Much of our current generic understanding of active systems has been gained by from the study of simple model systems, 
such as the Vicsek  \cite{Vicsek1995} and Active Brownian Particles (ABP) \cite{Fily2012, CatesEPL, Lino2018} models.  
In particular, the Vicsek model describes self-propelled agents with a tendency to align their velocities with their neighbors, predicting the emergence of flocking, a synchronized state where particles move collectively along a given direction  \cite{Vicsek2012}. 


The exchange of information in populations of mobile agents can be represented by the introduction of extra internal degrees of freedom in active matter models. 
Local interactions between agents involving such internal degrees of freedom are known to give rise to emergent phenomena
such as synchronization, epidemic spreading or social consensus.
While the emergence of synchronized states has been traditionally studied within the framework of coupled oscillators sitting on static networks \cite{AcebronRev, ArenasRev}, many examples of synchronization phenomena have been found in systems of (locally coupled) mobile agents,  such as genetic oscillators  \cite{danino2010synchronized} or people crowds \cite{strogatz2005}. Consistently, several authors have recently considered the impact of motility on  synchronization \cite{frasca2008synchronization,  Peruani2010, Fujiwara2011, Uriu2014, PhysRevX.7.011028}, a question which is also attracting increasing attention  in the context of epidemiology and computational social sciences. Agents carrying an infectious disease may transmit it, and their motion trigger the physical interactions responsible for the contagion process \cite{frasca2006dynamical, buscarino2008disease, buscarino2014local}. 
Mobile interacting agents have been used to model the human dynamics of face-to-face interactions in social gatherings \cite{PhysRevLett.110.168701} or via mobile communication devices \cite{onnela2007}, as well as to investigate the emergence of cooperative behaviors \cite{PhysRevE.79.067101} and consensus \cite{olfati2007consensus, PhysRevE.85.016113} among individuals.

 In all the above mentioned studies, however, the spatial location of the agents evolves independently of their internal state (e.g. phase, opinion, epidemic state, etc.).
 Only very recently, a few works started considering a feedback between the agent's internal states and the way they move in space.
 Coupling the tendency to synchronize of globally coupled mobile phase oscillators with their spatial attraction produces novel self-sustained structures \cite{OKeeffe:2017aa},
 while identifying the internal phase with the self-propulsion velocity of locally interacting active oscillators induces qualitatively new synchronization phenomena, such as mutual flocking and chiral sorting \cite{levis2018sync}.
 Alternatively, the internal variables of the agents  may represent not phases but opinions, whose dynamics depends on the spatial location of the agents which, in turn, is affected by their local social interactions \cite{Starnini:2016aa}. 
 Such feedback between mobility and social dynamics may lead to the emergence of metapopulation structures, formed by groups of individuals sharing similar opinions,  i.e. echo chambers \cite{garrett2009echo}.
Therefore, introducing a  feedback between the dynamics of an internal degree of freedom (phases, opinions) and the agents' motion in physical space has a dramatic impact on their collective behavior, for coordination in both real space (e.g. collective motion, pattern formation) and in the abstract space of the internal states (e.g. phase synchronization, consensus).

In this paper, we propose a new reference model that introduces a feedback between the agents'€™ mobility and an internal state, characterizing the spreading of information across the population.
 We consider self-propelled agents which align (or synchronize) their direction of motion, following a local interaction based on the Kuramoto model, a paradigmatic model of phase synchronization  \cite{kuramotoBook, AcebronRev}.
On top of that, agents carry an internal phase subjected to the dynamics of a Susceptible-Infected-Susceptible (SIS) process \cite{siroriginal}, characterizing  information spreading.  Individuals aware of the information (infected) transmit their internal phase to unaware (susceptible) neighboring individuals, fueling social contagion.
 The internal phase of an agent represents its tendency to move along a given direction, and thus directly influences its motion.
We show that motility can enhance the spreading dynamics across a population and, conversely, the diffusion of information can induce a variety of cooperative states in real space, such as flocking, swarming and pattern formation. Such rich and novel phenomenology crucially hinges on the feedback between the agents' mobility and their internal state.
 
The article is structured as follows: in Section \ref{sec:model} we introduce the model. Section \ref{sec:VicsekSIS} is devoted to explore the consequences of the feedback between spreading dynamics and motility, within the phase space of the parameters governing the Kuramoto and SIS interactions.  Then, starting from Section  \ref{sec:SISFlocking}, we focus on the special case in which the Kuramoto-like coupling is removed, describing how the epidemic process alone is able to induce flocking. In section  \ref{sec:Patterns} we discuss the nature of the structures which characterize the emergence of order in the system.  A final discussion and concluding remarks are reported in Section \ref{sec:Conclusion}.

\section{The Model}\label{sec:model}

We consider a system made of  $N$ self-propelled (point-like) agents moving in a $2d$ volume  $V=L_x\times L_y$ with periodic boundary conditions. At time $t$, particles are located at $\boldsymbol{r}_i^t=(x^t_i,\,y^t_i)$. They are self-propelled along their orientation $\boldsymbol{p}_i^t=(\cos \theta^t_i,\,\sin \theta^t_i)$ with a constant velocity $v_0$. 
The spatial evolution of the agents is simply given by
\begin{equation}
\label{eq:r_ev} 
\boldsymbol{r}_i^{t+dt}=\boldsymbol{r}_i^t+v_0\,(\cos\theta_i^t,\sin\theta_i^t) \, dt.
\end{equation}

The agents are subjected to a Susceptible-Infected-Susceptible (SIS) epidemic process. 
Each agent $i$ is endowed with an internal binary variable $s_i^t = \{ 0,1 \}$, representing its epidemic state, susceptible ($s_i=0$) or infected ($s_i=1$).  
At each time step, infected agents decay spontaneously to the susceptible state with probability $\mu$,  
 while susceptible agents may become infected upon contact with infected neighbors, with probability $\lambda$. 
 Two agents $i$ and $j$ are considered neighbors if $|\boldsymbol{r}_i-\boldsymbol{r}_j|\leq R$, the coupling range. 
The dynamics of the  SIS variables can  thus be cast by the  reactions
 \begin{equation}
 S+I \xrightarrow{\lambda} 2I \,, \, \ I \xrightarrow{\mu} S
  \end{equation}
If a susceptible agent is surrounded by more than one infectious neighbor, each connection hosts statistically independent stochastic infection processes.   

\begin{figure}[tbp]
    \includegraphics[width=8.4cm,angle=0]{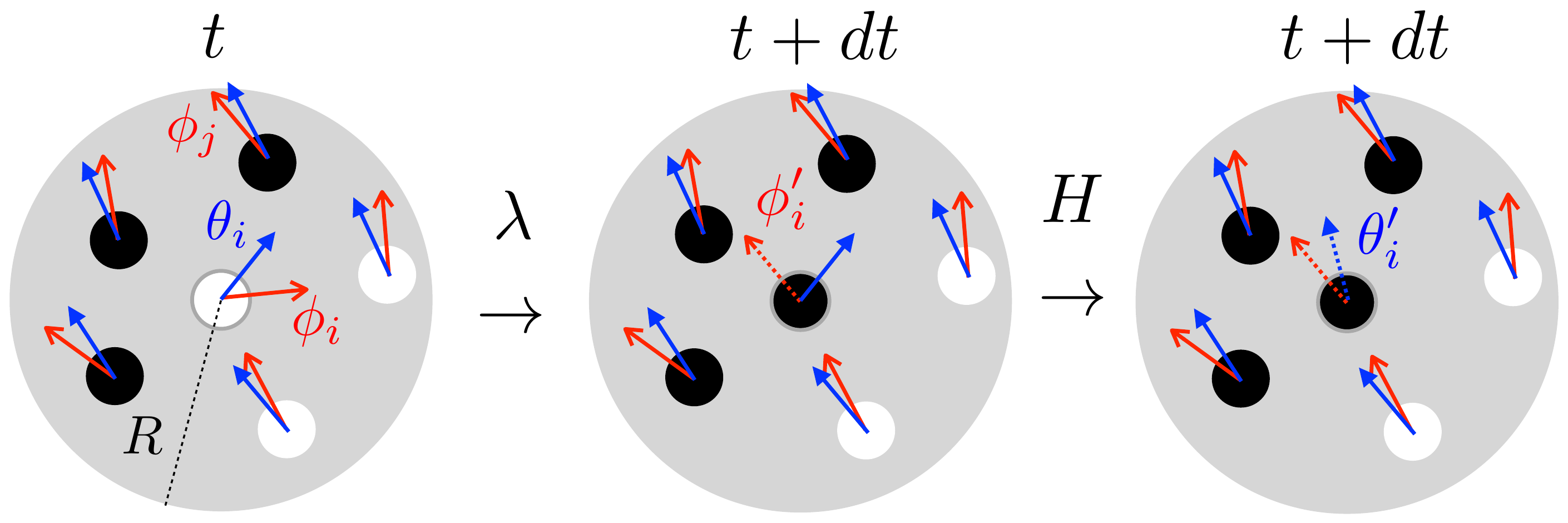}
          \caption{Illustration of the velocity-alignment interaction mechanism mediated by the SIS process. 
          Susceptible (infected) agents are represented in white (black).
           For each agent $i$, the internal phase $\phi_i$ (orientation $\theta_i$) is indicated by a red (blue) arrow. 
          In the first illustration (left), at time $t$, a susceptible agent $i$, at the center of the grey area, is interacting with other agents located within a distance $R$. 
          In the second illustration, at time $t+dt$, agent $i$ is infected by agent $j$ (with probability $\lambda$) and adopts its internal phase $\phi_j^t$. 
         Then, agent $i$ aligns its orientation $\theta_i^{t+dt}$ with its new internal phase $\phi_i^{t+dt}=\phi_j^t$ with a probability $H$ (third illustration).    }
       \label{fig:cartoon}
\end{figure}

Furthermore, each agent $i$ carries an internal phase $\phi_i \in [0, 2\pi]$ which evolution is dictated by  the SIS process. 
At a given time $t$, if an agent $i$ is infected by another neighboring agent $j$, agent $i$ adopts the internal phase of agent $j$, $\phi_i^{t+dt} = \phi_j^t$ (see cartoon Fig. \ref{fig:cartoon}). 
If an infected agent $i$ becomes susceptible, he adopts a new phase, randomly extracted from the uniform distribution $U(0, 2\pi)$.
The phase of susceptible agents evolves in time according to a Gaussian white noise, while the phase of infected agents is constant as long as they remain in the infectious state.  
The coupled SIS variables $\{ \phi_i, s_i \}$ of agent $i$ thus evolve as 

\begin{equation}
\{  \phi_i, s_i \}^{t+dt}   =
  \begin{cases}
  \{  \tilde{\phi_i}, 0 \}   &   \mu s_i^t   \\
   \{  \phi_i^t, 1 \}  & (1-\mu) s_i^t \\
 \{ \phi^t_{j\in\partial_i}, 1 \}  & \lambda (1-s_i^t) \\
 \{ \phi_i^t + \delta\phi_i\,dt, 0 \}  &   (1-\lambda) (1-s_i^t).
    \end{cases}
    \label{eq:SIS}
\end{equation}
The first possible update corresponds to the transition $I \rightarrow S$ of an infected agent ($s_i^t=1$) with probability $\mu$. In this case, the agent picks a new phase $\tilde{{\phi}_i}$ at random  from the uniform distribution $U(0, 2\pi)$. 
The second case corresponds to the complementary step, an infected agent remains in the same epidemic state $s_i^t=1$ and with the same phase $\phi_i^t$ with probability $(1-\mu)$. 
The third possible update corresponds to the transition $S + I \rightarrow 2I$, a susceptible agent $i$ ($s_i^t=0$) becoming infected upon contact with a neighbor $j$ with a probability  $\lambda$. The neighborhood of $i$, denoted $\partial_i$, is defined by $R$.  The effect of such infection is the transmission of the internal phase from $j$ to $i$.
The last case corresponds to a susceptible agent ($s_i^t=0$) remaining in the same epidemic state and its internal phase performs a free diffusive step: $\phi_i^{t+dt}=\phi_i^{t} + \delta\phi_i^t\, dt$, where $\delta\phi$ is a Gaussian white noise term verifying
\begin{equation}
\label{eq:NoiseA} 
\langle{\delta\phi}_i^t \, {\delta\phi}_j^{t'}\rangle = 2D_{0}\delta_{t t^{\prime}} \delta_{ij}\ \,,\ \langle{\delta\phi}_i^t\rangle=0\,.
\end{equation}
The parameter $D_{0}$ sets the noise strength.
The evolution of the internal phase $\phi$ is thus slaved to the SIS dynamics: if few agents are infected, then the internal phases will be uniformly distributed in the population, while the more agents are infected, the more agents will share the same internal phase. 

Finally, the SIS dynamics is coupled with the mobility of the agents through their orientation $\theta$, which sets their direction of self-propulsion in the 2$d$ plane, see Eq. \eqref{eq:r_ev}. 
The dynamics of the orientation of agent $i$ is based on the Kuramoto model of phase oscillators \cite{AcebronRev}, 
\begin{equation}
\label{eq:theta_ev} 
 \theta_i^{t+dt}-\theta_i^t=\left[K\sum_{j\in\partial_i} \sin(\theta_j^t-\theta_i^t)+H \sin(\phi_i^t-\theta_i^t)\right] \, dt .
\end{equation}
The first term on the right-hand side corresponds to a Kuramoto  interaction of strength $K\geq0$, which tends to align the self-propulsion direction of neighboring agents. 
Since $\theta_i$ dictates the direction of  self-propulsion of the agents,  such ferromagnetic interaction term is akin to the  Vicsek model \cite{Vicsek1995, Chate2008, Vicsek2012} and can be equivalently seen as a  continuous-time variant of the original agent-based model \cite{Chepizhko2010, Farrell2012, Liebchen2017, Martin2018}. 
The crucial new feature of the present model is the coupling between the internal phase $\phi_i$, governed by the SIS process, and the self-propulsion direction of the agents $\theta_i$. Such back-coupling is made explicit by the second term on the right-hand side Eq. \eqref{eq:theta_ev}, which accounts for the tendency of each agent to adjust its velocity with its internal phase with a rate $H\geq0$. The larger $H$, the more efficiently, or faster, each agent $i$ will align his orientation $\theta_i$ with its internal phase $\phi_i$.  Fig. \ref{fig:cartoon} illustrates this mechanism, which is mediated by the infection process.
Note that the Kurmoto and SIS interactions, controlled by $K$ and $H$, respectively, can act constructively or not, depending on the degree of alignment between the internal phase and the orientation of the infected neighbors.

In summary, our model is built on three main ingredients: i) Self-propelled agents, whose motion is characterized by velocity $v_0$ and orientation  $\{ \theta \}$;   ii) a Kuramoto-like interaction controlled by the coupling constant $K$, which tends to align the self-propulsion direction of neighboring agents; and iii) an SIS epidemic process, 
in which infected agents transmit their internal phase $\{ \phi\}$ to susceptible ones, while all agents tend to align the self-propulsion direction $\{\theta\}$ with their internal phase $\{\phi\}$. 
Our model introduces a feedback mechanism between the agents' mobility and the epidemic process, in which the SIS dynamics affects the angular directions of the agents through the orientation-phase coupling term of strength $H$ in Eq. \eqref{eq:theta_ev}, and, in turn, the infection process depends on the spatial position of the agents, since it occurs only between close enough agents.
The micro-state of the system is characterized by the set of variables $\Gamma^t=\{ \boldsymbol{r}^t,\theta^t,\phi^t, s^t \}_i$, while the control  parameters of the model are: the self-propulsion velocity $v_0$, the average density $\rho_0=N/L^2$, the SIS epidemic parameters $\lambda$ and $\mu$, the Kuramoto interaction strength $K$,  the phase-orientation coupling $H$, the noise intensity $D_{0}$ and the interaction range $R$.


In order to establish the collective behavior of the model on general grounds, we need to reduce the dimensionality of the parameter space and identify the most relevant control parameters. We set the ratio between infection and recovery probabilities of the SIS process, as usual, by setting $\mu=1$. This choice is equivalent to set the time scale of the epidemic process. 
We then keep $\rho_0=10$, $D_{0}=5.10^{-4}$ and $R=1$ fixed (setting the units of time and length), while we systematically explore the behavior of the model in the four-dimensional parameter space $\{\lambda , K, H, v_0 \}$.
 In order to identify the relevance of finite-size effects, we  simulate systems of $N = 10^3$ up to $3.10^4$ agents, both in square $L_x = L_y = L$ and slab $L_x = 6L_y$ geometries. Slab geometry is used to properly characterize the structures, or patterns, emerging in the system (see Section \ref{sec:patterns}).
If not stated differently (see sec. \ref{sec:VicsekSIS}), we initialize the system in a disordered state: each agent $i$ has a randomly selected position $\boldsymbol{r}_i$, orientation $\theta_i$ and internal phase $\phi_i$. 
All agents are initially in a susceptible state, except for a set of $fN$ randomly chosen agents, with $f = 0.1$, set to the infectious state. 
We numerically integrate the dynamics of the model Eqs. \eqref{eq:r_ev}, \eqref{eq:SIS} and \eqref{eq:theta_ev} by using a discrete time-step $dt=0.1$, until a steady state is reached, characterized by a steady value of global quantities like the polarization and the fraction of infected agents (see  Section \ref{sec:VicsekSIS}). 
 We consider a modified SIS process which eliminates the absorbing state with no infected agents, obtained by reinfecting the last remaining infected agent immediately after it is cured. The steady state of the modified SIS process has been shown to be equivalent to the metastable state of the standard SIS process \cite{PVM_MSIS_star_PRE2012}.

\begin{figure*}
    \includegraphics[width=16.8cm]{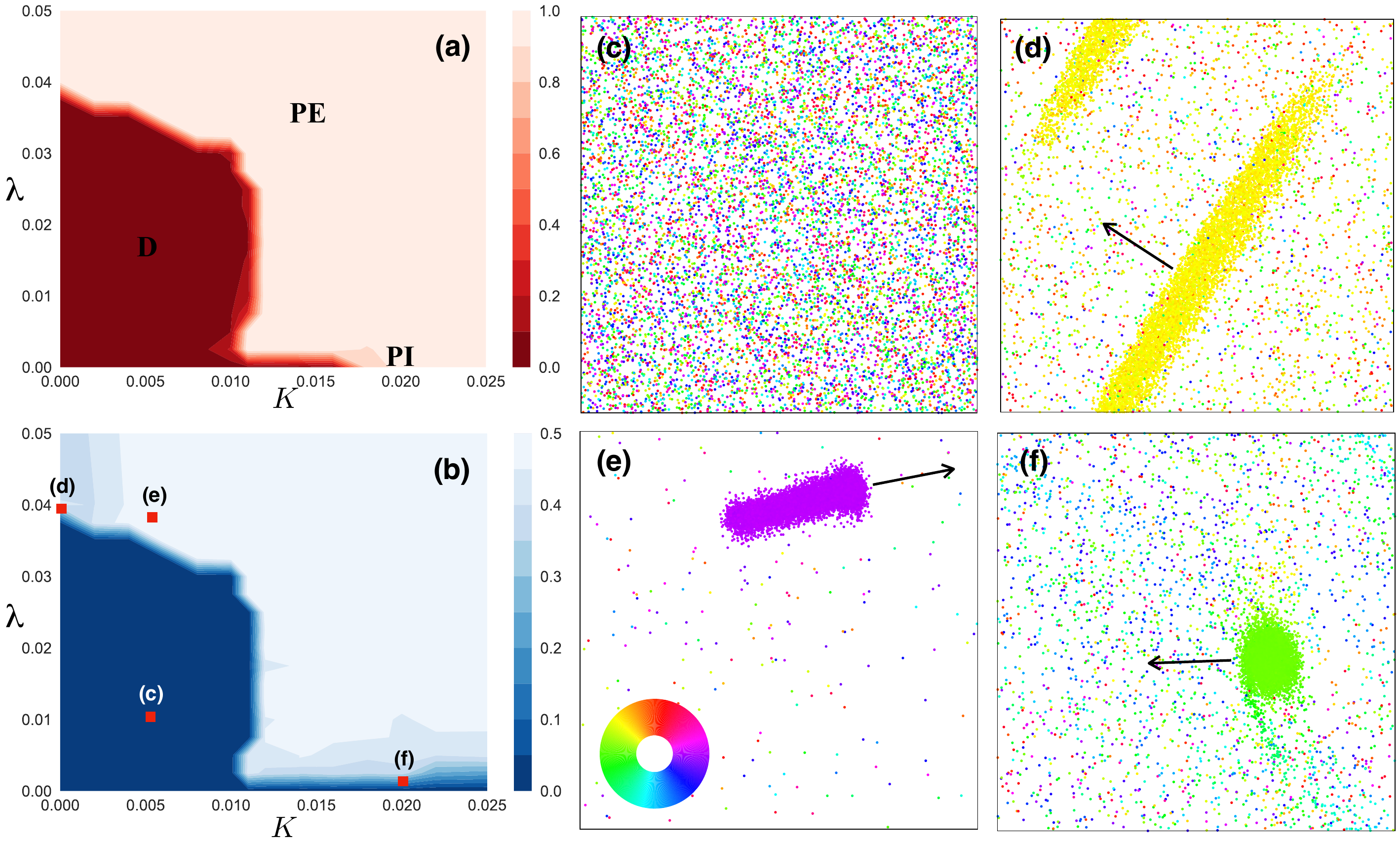}
      \caption{ \textbf{Phase diagrams in the ($K$-$\lambda$) plane.} Average synchronization $\av{Z}$  (a) and the average prevalence $\av{n}$(b) in the ($K$-$\lambda$) plane at fixed  $H=0.1$, $v_0=1$. Note that in the epidemic phase the prevalence does not reach one, $\rho \simeq 0.5$, because we choose a high recovery probability, $\mu=1$. Symbols denote the parameters corresponding to the snapshots shown in (c).
      Representative snapshots illustrating the different regimes (c): for ($K$-$\lambda$)=$(0.005,\, 0.01)$ (disordered state) (i) , $(0,\, 0.04)$ (polar order induced by information spreading) (ii), $(0.005,\, 0.04)$ (polar order induced by both information spreading and velocity-alignment) (iii) and $(0.02,\, 0)$ (order induced by the velocity-alignment only) (iv). In these snapshots (as well as in all the other snapshots shown throughout the article) agents are colored accordingly to their orientation (an arrow showing the orientation of the swarms has been included for clarity).  
      Here we used $N=10^4$ and a disordered initial condition. 
       \label{fig:maps_K_l}}
\end{figure*}

\section{Interplay between flocking and spreading dynamics}\label{sec:VicsekSIS}

Our model captures two local interaction mechanisms: the Kuramoto interaction, controlled by the coupling strength $K$, and the epidemic spreading, whose virulence is quantified by the infection probability $\lambda$. 
The former favors the emergence of synchronization, or flocking, while the second the diffusion of a given information among the population. 
The emergence of  flocking can be identified by the polarization
$$Z(t) \equiv N^{-1} \left| \sum_i e^{i \theta_i(t)}\right|\, $$
which here, because the self-propulsion direction if associated to a phase variable,  is  equivalent to the Kuramoto order parameter of phase synchronization \cite{AcebronRev}. 
The prevalence $n(t)$ of the epidemic process, defined as the fraction of infected agents at time $t$,
 $$n(t) \equiv N^{-1} \sum_i s_i(t) \,$$
characterizes  the epidemic state \cite{PastorRev}.

We first explore the steady states of the model for systems defined on a square   box  at fixed $H=0.1$ and $v_0=1$, while varying $\lambda$ and $K$. 
Fig. \ref{fig:maps_K_l} shows the polarization $\av{Z}$ (a) and the prevalence $\av{n}$ (b) in the phase space $\lambda - K$. [The brackets $\av{*}$ denote an average over steady-state configurations.] 
The first remarkable feature of Fig. \ref{fig:maps_K_l} is the similarity between the phase diagrams of the polarization $\av{Z}$  and the prevalence $\av{n}$. 
This means that the synchronization and epidemic transitions are characterized by similar critical boundaries $(K_c, \lambda_c)$ in the $\lambda - K$ plane. 
As expected, for low enough values of $\lambda$ and $K$, the system is disordered (dark areas of plots a) and b) of Fig. \ref{fig:maps_K_l}), with a negligible fraction of infected agents and a global polarization close to zero. 
In this disordered phase, agents move randomly (see  Appendix A for more details on how they move in the dilute $\lambda\to0$ limit) and most of them are in the susceptible state.  As the parameters $\lambda$ and $K$ increase, the system can exhibit the emergence of macroscopic ordering: both flocking and epidemic outbreak (light areas of the plots). 
On the one hand, the Kuramoto coupling of strength $K$ is able to trigger flocking: above some threshold value $K_c$, the system breaks  rotational symmetry and acquires a spontaneous net global polarization. 
On the other hand, the infection probability $\lambda$ triggers an epidemic transition: above some critical value $\lambda_c$, a macroscopic fraction of the system is infected.  

The interplay between the SIS and the Kuramoto dynamics enhances both flocking and information diffusion, which are achieved in regions of the phase space where the Kuramoto interaction and the epidemic spreading separately do not trigger nor a polar neither an endemic state. 
 Strikingly though, even in the absence of velocity alignment (i.e. $K=0$), the SIS process is able to spontaneously break the rotational symmetry of $\theta$ and generate polar order (or flocking), as shown in Fig. \ref{fig:maps_K_l}. Such behavior shows that the feedback between the SIS process and the motility of the agents drives a flocking phase transition in the absence of explicit velocity-alignment interaction. 
The epidemic threshold is $\lambda_c \simeq 0.04$ for  $K=0$ and  decreases as $K$ increases. 
The onset synchronization at $\lambda=0$ occurs at $K_c \simeq 0.01$. For values of $K>K_c$, $\lambda_c$ remains almost constant and close to $\lambda_c \simeq 0.003$. 
However, the epidemic and synchronization boundaries are different in the region $K > 0.01$ and $\lambda \lesssim 0.003$. 
We can thus differentiate three regimes, or phases, summarized in Fig. \ref{fig:maps_K_l}: (i) the disordered phase (D) for which $Z\approx n\approx 0$; (ii) the \emph{polar endemic phase} (PE) for which $Z>0$ and $n> 0$; and (iii) the polar inactive phase (PI) for which $Z>0$ and $n\approx 0$.  In the PI regime, the prevalence is zero (as expected), yet the system  sustains polar order.

Besides the emergence of flocking, 
the establishment of polar order  is accompanied by the spontaneous formation of \emph{swarms}, dense structures showing orientational order and thus  moving  coherently. In Fig. \ref{fig:maps_K_l} we show several characteristic  snapshots of the system in  different regions of the  $\lambda - K$ plane.  Agents accumulate into traveling structures reminiscent of what is  generically found in systems of aligning self-propelled particles, and responsible of the first-order like nature of the flocking transition \cite{Chate2008, Solon2015}. 
In particular, we observe the formation of bands when flocking is uniquely triggered by the epidemic spreading, meaning, in the limit $K=0$ (see snapshot d). 
At finite values of $K$ and $\lambda$, as the system gets deeper into the endemic polar phase, such traveling structures become more compact and coherent (keeping their shape for very long periods of time). Such increase in the stability of the swarm is followed by a decrease of  the density of the surrounding incoherent gas  (see snapshot Fig. \ref{fig:maps_K_l} e). In the inactive polar phase, swarms are still present, but subjected to larger fluctuations than in the endemic polar phase (see snapshot Fig. \ref{fig:maps_K_l} f).

The emergence of bands in the case $K=0$ is particularly interesting, since we recover a phenomenology which reminds the behavior of Vicsek-like systems from a radically different perspective, i.e. as arising from the dynamical feedback between motility and information spreading. 
Moreover, from an epidemic perspective, the study of this limit case allows us to investigate the impact of mobility on the SIS model and, in particular, how the agents' velocity affects the epidemic threshold. 
Therefore, we will explore the $K=0$ limit case in details in the following, while other limiting regimes of the model, including a precise analysis of the motion of free agents ($K=\lambda=0$), are detailed in Appendix A.

\section{Epidemic process induces flocking}\label{sec:SISFlocking}
\label{sec:K0}

\begin{figure}[tbp]
\includegraphics[width=8.5cm]{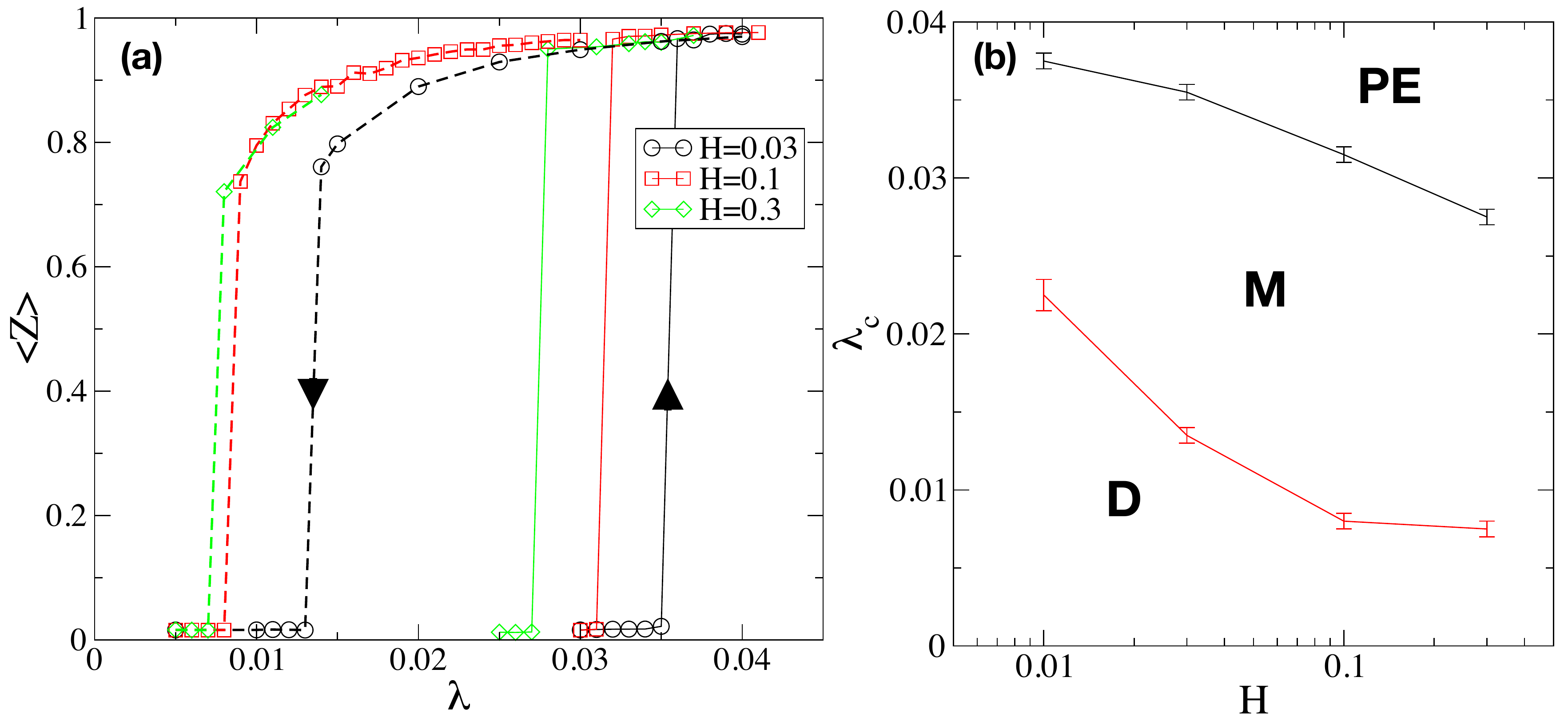}
\caption{ (a): Polar order parameter as a function of $\lambda$ for several values of $H$ shown in the key. The data points connected by continuous lines where obtained by letting relax the system after a high-$\lambda$ quench, while the ones connected by dotted lines refer to steady states obtained after a low-$\lambda$ quench. (b): Onset of polar ordering $\lambda_c^{\pm}$ obtained from the data shown in (a). Black and red points correspond to $\lambda_c^{\pm}$ from the high-$\lambda$ and low-$\lambda$ quench, respectively.
The polar endemic (PE), metastable (M) and disordered (D) regions are highlighted. Here we used  $v_0=0.1$ and $N=3.10^3$. }
\label{fig:varh}
\end{figure}

In this and the following Sections we explore in detail the behavior of the model when the Kuramoto interaction is removed ($K=0$) and the agents' motion is completely determined by the epidemic process. 
 In this limit, the epidemic probability $\lambda$  controls the emergence of flocking from a disordered state to an endemic polar one. 

Fig. \ref{fig:varh} (a)  shows  $\av{Z}$  as a function of  $\lambda$, for different values of $H$ at fixed $v_0=0.1$.  The data shown was obtained following two different preparation protocols: (i) starting from a completely disordered configuration and letting it evolve towards a steady-state (we call that a high-$\lambda$ quench); and (ii) starting from a endemic polar state and letting it relax (hence called a low-$\lambda$ quench).   
Starting from a disordered state and doing a high-$\lambda$ quench the order parameter jumps from $\av{Z}\approx 0$ to $\av{Z}\approx 1$ at a given value of $\lambda=\lambda_c^+$, which sets the limit of stability of the disordered state. Conversely, when letting the system relax after a low-$\lambda$ quench, the order parameter jumps from $\av{Z}\gtrsim 0.7$ to $\av{Z}\approx 0$ at  $\lambda=\lambda_c^-$, indicating the limit of stability of the ordered state. Typically, $\lambda_c^-<\lambda_c^+$, a behavior is consistent with a discontinuous, or first-order,  phase transition: hysteresis and abrupt changes of the order parameter around the transition. 
Following the first-order transition picture, the two instability lines $\lambda_c^\pm$, or spinodals, define a metastable region in the $\lambda-H$ plane, shown in Fig. \ref{fig:varh} (b). 

The epidemic thresholds $\lambda_c^\pm$ depend on the strength of the phase-orientation coupling $H$ (see  Fig. \ref{fig:varh}).
As $H$ increases, the agents adjust faster their direction of motion $\theta_i$ with their phase $\phi_i$, promoting the propagation of the information. The coordinated motion induced by the coupling between the agent's epidemic state and its orientation, enhances their interaction time as compared to agents which move independently of their internal epidemic state. Locally aligned agents have longer time to exchange information than non-aligned ones, thus fostering synchronization, or flocking, among them. Such positive feedback loop is at the origin of the rich phenomenology reported in this article.


 \begin{figure}
 \includegraphics[width=8.5cm]{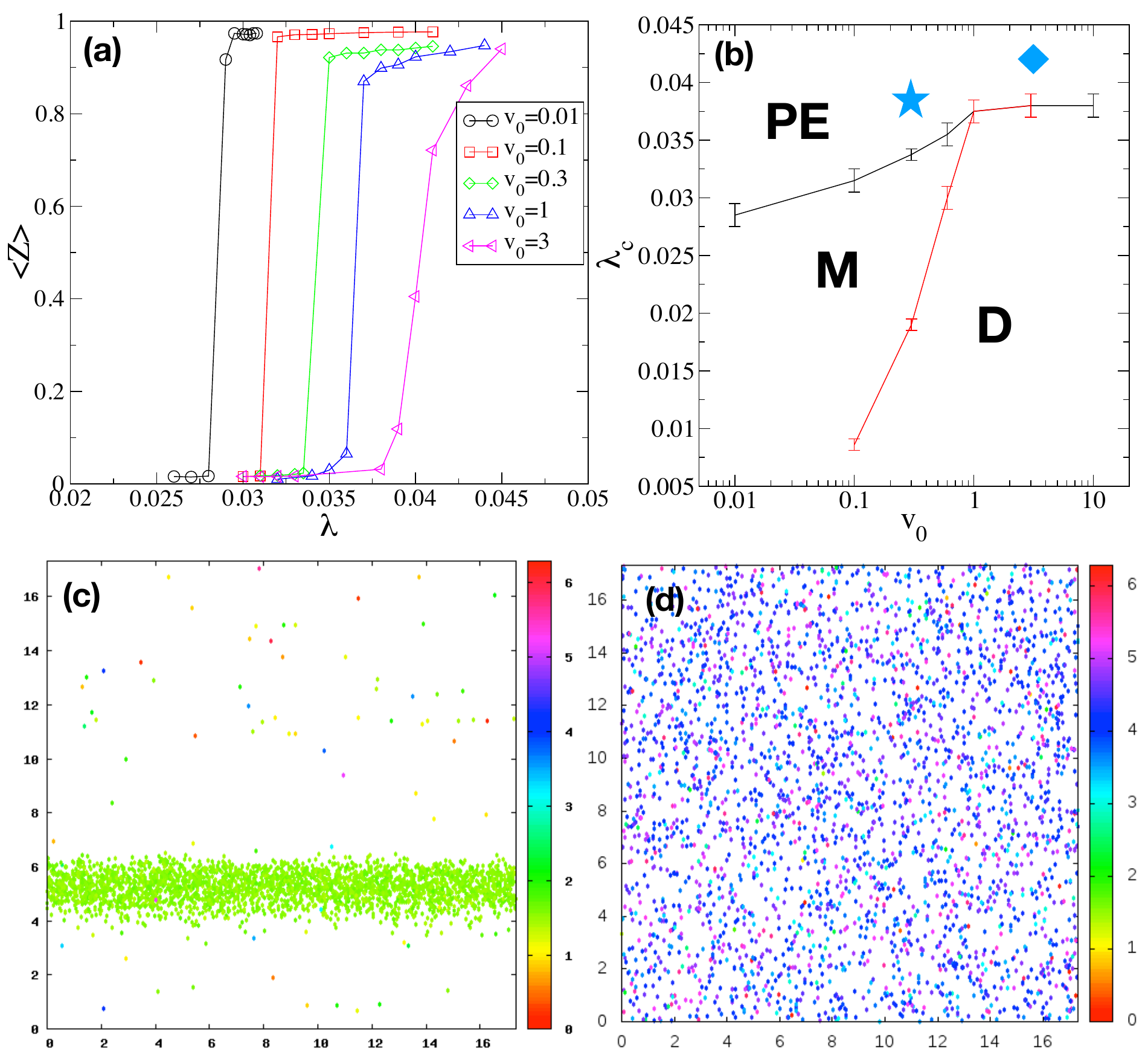}
\caption{ (a) Polar order parameter as a function of $\lambda$ for several values of $v_0$ shown in the key.  (b) Location of the onset of flocking obtained after letting the system relax from a disordered (plan line-points) and ordered (dotted line-points) initial configurations. The polar endemic (PE), metastable (M) and disordered (D) regions are highlighted.  (c) Typical long-time snapshots of the system  for $v=0.3$ and $\lambda=0.038$ (indicated by a star in plot b)) and (d) $v=3$ and $\lambda=0.042$ (indicated by a diamond in plot b). In all the cases $H=0.1$ and $N=3.10^3$. }
\label{fig:varV}
\end{figure}

To better understand the impact of motility on information spreading, we  now focus our attention on the role played by the agent's velocity. 
In Fig. \ref{fig:varV} (a) we show  $\av{Z}$ as a function of  $\lambda$, for different values of $v_0$ at fixed $H=0.1$. 
This data was obtained  using a high-$\lambda$ quench. 
For $v_0\lesssim 1$, we observe the sharp increase of the order parameter $\av{Z}$ and hysteresis around the transition,  supporting the first-order scenario  discussed above. 
As $v_0$ increases from $v_0=10^{-2}$ to $v_0=3$ the onset of flocking shifts to larger values of $\lambda$.
Interestingly,  the evolution of $\av{Z}$ becomes smoother when $v_0 \gtrsim 1$. In order to perform a systematic analysis of the phase behavior, we identify the onset of flocking by $\av{Z}(\lambda_c)=0.05$ and report the results in Fig. \ref{fig:varV} (b), both using the high-$\lambda$ and  low-$\lambda$ quench procedure. 
The  onset of flocking $\lambda_c^{\pm}$ using both kind of quenches saturates to a well defined value at large self-propulsion velocities. The metastability region between $\lambda_c^-$ and $\lambda_c^+$ shrinks as $v_0$ increases, and it eventually  disappears for $v_0 \gtrsim 1$. 
It is worth to notice here that the static  limit is singular and our results cannot be extrapolated to $v_0\to 0$ (a precise study of this limit in the context of the Vicsek model was carried in \cite{Baglietto2009computer}). 
The static version of our model corresponds to a 2$d$ XY model coupled with a SIS process defined on a simply connected random geometric network.

The absence of hysteresis and the smoothing of the polarization curves for large velocities suggests that the nature of the flocking transition changes from discontinuous to continuous for  $v_0 \gtrsim 1$. This is further supported by the structure of the system. We show in Fig. \ref{fig:varV} (c) and (d) typical  snapshots of the system obtained in the small ($v_0<1$) and high velocity ($v_0 \gtrsim 1$) regimes above  $\lambda_c^+$ .  At low velocities the system de-mixes into a high-density swarm, in the form of a traveling band,  and a low density incoherent background. At larger velocities, the ordered phase consist on a homogenous flock and the characteristic density patterns of Vicsek-like models is never observed. Such change in morphology shows that the ordering mechanism in the fast and slow velocity regimes is qualitatively different, and therefore the very nature of the flocking phase transition that takes place. 
The emergence of traveling bands is known to be responsible of the discontinuous nature of the flocking transition in Vicsek-like models \cite{gregoire2004,Chate2008}. However,  the amplitude of the self-propulsion velocity $v_0$ is not expected to qualitatively affect the emergence of flocking  at large scales, as long as $v_0>0$. 
This in sharp contrast with our results: here, the value of $v_0$ directly affects how the agents exchange their information. 

At low velocities, neighboring agents interact with each other for long periods of time,  favoring information spreading at short distances. Such information spreading leads to their coordinated motion. 
Two oriented agents move together through space, enhancing their interaction time and thus fostering information spreading. This positive feedback loop is at the origin of the enhancement of flocking and information spreading due to the presence of dense swarms. The slower the particles are, the denser are the swarms and the more efficient is the local information spreading process. As the agents velocity increases, their mutual interaction time decreases, and the efficiency of the local information spreading is reduced. In the limit of large velocities, the motion of the agents becomes faster than the time needed for the information to propagate. In this limit, the spatial structure of the system is lost,  the evolution of the agents and the propagation of their internal state decouple and thus no local structures are  possible.  This corresponds to a mean-filed like limit where agents can  be considered to interact globally with all the agents in the system,  as illustrated by the snapshot  Fig. \ref{fig:varV} (d). 

From the point of view of the epidemic process, it is interesting to highlight such feedback mechanism.  
In previous epidemic models of mobile agents that do not couple  the SIS dynamics and the agents' mobility, the epidemic threshold  increases as the velocity decreases. That is, a system of slowly moving agents reaches the endemic state for larger values of $\lambda$ than fast moving ones, and the smallest epidemic threshold is reached in the limit of homogeneous mixing \citep{buscarino2008disease}. In the present case, the emergence of flocking associated with the epidemic process generates dense swarms which, in turn, foster epidemic spreading.

\begin{figure}
  \begin{center}  
    \includegraphics[width=9cm,angle=0]{\FigPath/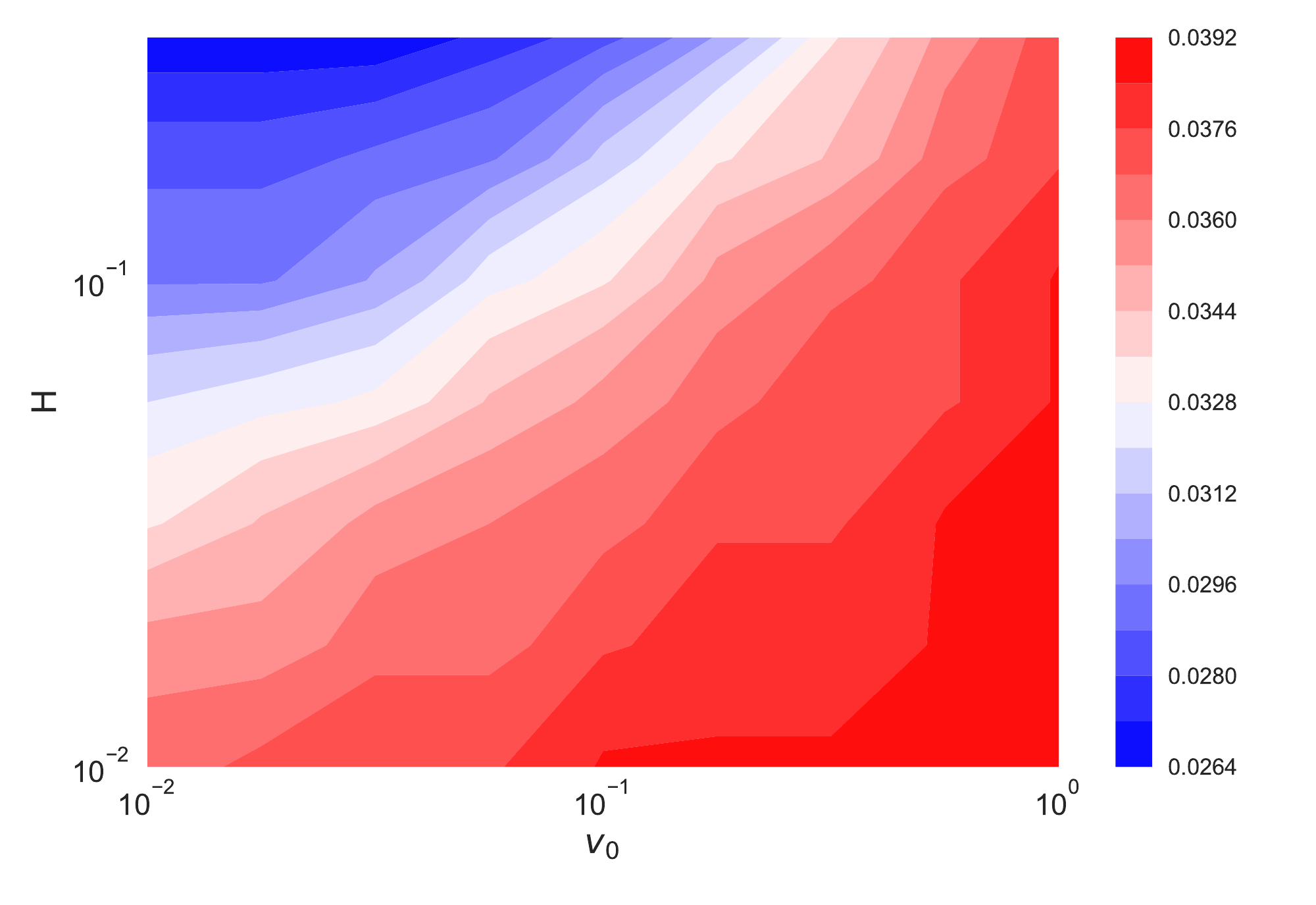}
      \caption{ Contour lines associated to the limit of stability of the disordered state  $\lambda_c^+$  in the $(H - v_0)$ plane (in log-log scale).  As made apparent in the figure,  the onset of flocking depends in a complex manner on both $H$ and $v_0$. \label{fig:map_lc}}
      \end{center}
\end{figure}

Altogether, we have obtained a clear picture of the impact of the rate of adjustment of the polarity of the agents with their internal state, $H$, and the velocity of the agents, $v_0$, on  information spreading and, equivalently, the emergence of flocking. Our results are summarized in Fig.  \ref{fig:map_lc}, where we show the limit of stability of the disordered state  $\lambda_c^+$ as a function of both $H$ and $v_0$. The contour lines do not follow any simple  linear or algebraic dependence on $H$ and $v_0$. This means that the interplay between the SIS dynamics and the motion of the agents in real space gives raise to truly complex behavior: the dependence of $\lambda_c^+$  on  $H$ and $v_0$  cannot be rescaled into a single variable. 

\section{Phase separation and pattern formation}\label{sec:Patterns}
\label{sec:patterns}

The observation of bands, hysteresis and order parameter jumps in the limiting case $K=0$ of our model is reminiscent of the Vicsek model phenomenology, although in a different parameter space and driven by an epidemic process. 
In this Section we show that the nature of the structures arising close to the flocking transition in our model departs from the one of the Vicsek model at a fundamental level.

\begin{figure}
  \begin{center}  
    \includegraphics[scale=0.35, angle=0]{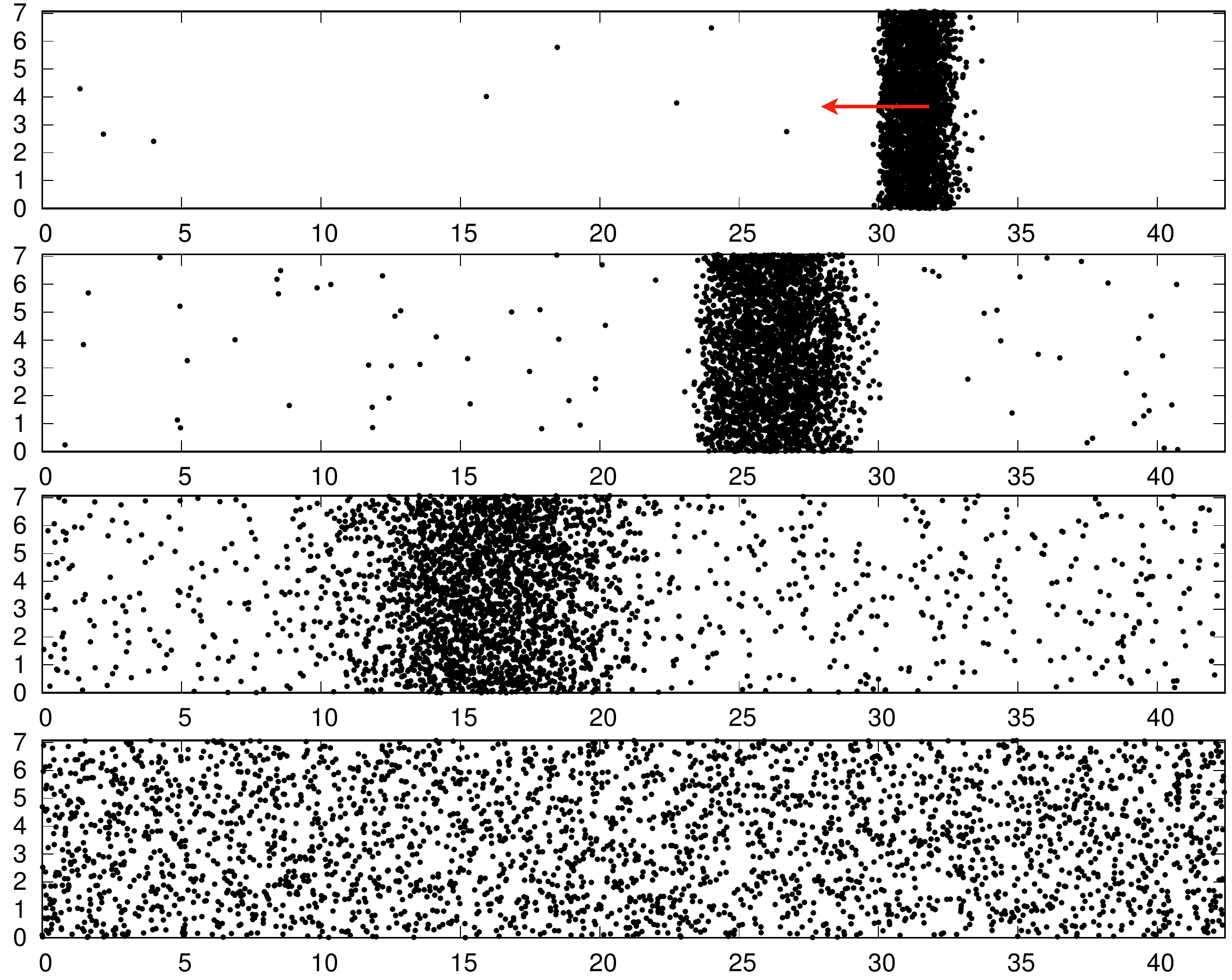}    
      \caption{Snapshots for (from top to bottom) $v_0=0.1$, $0.3$, $1$, $3$ in the slab geometry $L_x=6L_y$. The red arrow indicates the direction of motion of the band.  In all cases $N=3.10^3$, $H=0.1$ and $\lambda-\lambda_c^+=10^{-3}$.   \label{fig:bands}}
      \end{center}
\end{figure}

The nature of the flocking phase transition in the original model proposed by Vicsek and collaborators \cite{Vicsek1995} has been intensively debated over the last two decades \cite{Aldana2009, Baglietto2009, gregoire2004, Nagy2007, Chate2008, Solon2015}. 
For systems with periodic boundary conditions, the formation of bands which are either parallel or perpendicular to the boundaries of the simulation box  are responsible for the discontinuous character of the flocking transition. 
In order to clarify the nature of the transition in our model, we should thus characterize such high-density bands. 
Starting from random initial conditions, we let the agents evolve on a rectangular $L_x\times L_y$ plane with periodic boundary conditions. We fix $L_x=6L_y$ to bias bands to wrap around the short side of the box. 
As shown in Fig. \ref{fig:bands}, such geometry stabilizes bands (in the low velocity regime) traveling along the horizontal axis. 
At low self-propulsion velocities such percolating structures are hard to reach in practice. 
Thus, because of the absence of bands  in the Vicsek model  at low values of $v_0$, the flocking transition might appear as continuous \cite{Aldana2009}. 
This can be understood  from the analysis of the associated continuum theory, which predicts that the onset of flocking is independent of $v_0$ but the growth rate of the associated long-wavelength instability of the homogeneous disordered state increases with $v_0$ \cite{Bertin2009, Mishra2010}. 
At larger values of $v_0$ the instability grows faster and it is thus easier to see it in simulations.

Phase coexistence in the Vicsek model is quite peculiar: it features micro-phase separation instead of full phase separation as in the liquid-gas scenario,  with a critical point at $\rho\to\infty$ \cite{Solon2015}. 
 Contrary to the expectations from conventional liquid-gas separation, the system de-mixes into a set of regularly spaced bands in coexistence with a gaseous background at a well defined density, set by the control parameters of the model. 
 In this scenario, increasing the average density of the system at fixed noise strength in the coexistence region does not change the size of the band(s), but rather their number. 
The nature of phase coexistence in our model is in contrast with this scenario. 
Interestingly, the formation of traveling bands induced by the SIS process is closer to the standard  liquid-gas coexistence. 
In Fig.  \ref{fig:bands} we show representative snapshots of the system at a fixed distance to the onset of flocking $\lambda-\lambda_c^+>0$, for different  values of $v_0$. 
As $v_0$ increases, the size of the bands grows until they eventually fill the whole simulation box. 
Although the width of the bands grow with $v_0$, their density decreases correspondingly. 
Such decrease in the density of the high-density phase is compensated by an increase of the density of the low-density phase, the surrounding gas, as expected from liquid-gas phase separation as the critical point is approached. 

\begin{figure}
  \begin{center}  
    \includegraphics[scale=0.3,angle=0]{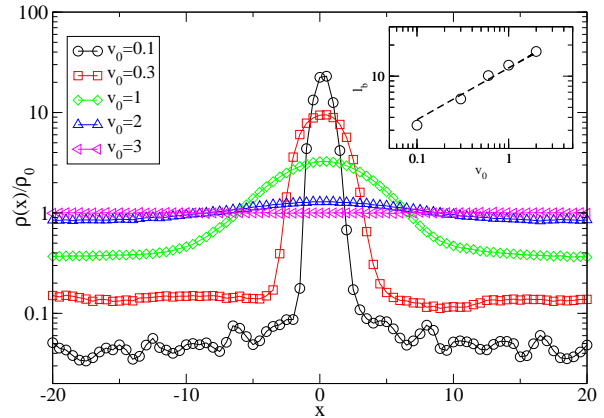}
    \includegraphics[scale=0.3,angle=0]{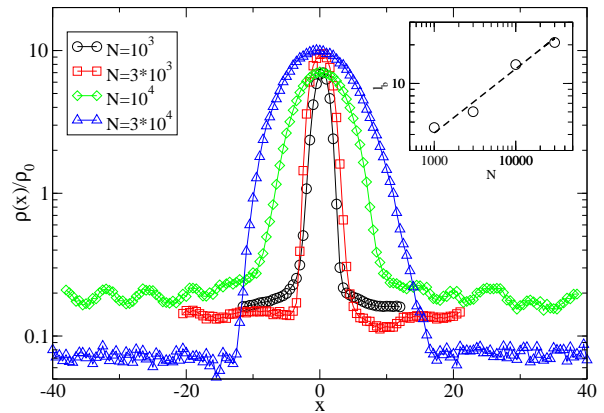}
          \caption{Density profiles averaged over the transverse direction. The $x$-axis corresponds to the location of the particles with respect to the center of mass of the system. Densities have been normalized by the average density of the system.  Top: Profiles corresponding to the situations illustrated in Fig. \ref{fig:bands}, i.e. for different values of $v_0$ at fixed $N=3.10^3$. The inset shows the band width $l_b$ as a function of $v_0$ (the dotted line corresponds to $l_b\propto \sqrt{v_0}$. Bottom: Density profiles at fixed $v_0=0.3$, yet varying the size of the system $N=10^3$,.., $3.10^4$. The dependence of the size of the bands  on $N$ is shown in the inset. As is made explicit by comparison with a $l_b\propto\sqrt{N}$ growth, the width of the bands scales with the linear size of the system.  In all cases  $H=0.1$.
      \label{fig:coex}}
      \end{center}
\end{figure}

To put the analogy with liquid-gas de-mixing into test, we analyzed the density profiles along the long-axis of the simulation box, shown in Fig. \ref{fig:coex}. 
These profiles where obtained by shifting the position of the particles into the reference frame of the center of mass and averaging over $10^4-10^5$ steady-state configurations. 
As shown in Fig. \ref{fig:coex} (a), at a given value of $v_0$, the system selects two coexisting densities: the density of the polar dense liquid $\rho_l$ and the gas $\rho_g$.  
In our simulations we never observed the formation of several bands. 
One could argue this might be due to the finite size of the system. 
However, as  the size of the system $N$ increases, also the width of the bands, $l_b$, does, while their mean density, $\rho_b$, remains almost identical (see Fig. \ref{fig:coex} (b)).
 This is the signature of \emph{macrophase} separation: the size of the dense phase scales with the size of the system.  
 Overall, our results provide a consistent picture about the role of $v_0$ in the establishment of order in the system and  the control of swarm formation. 
 The smaller the velocity of the agents, the  denser the swarms generated, fostering information spreading and therefore shifting the onset of flocking to lower values of the information spreading rate.

\section{Discussion and future perspectives}\label{sec:Conclusion}

The present work unveils the effects of a positive feedback between self-propulsion and information spreading in populations of mobile agents. Such feedback is responsible for the enhancement of information spreading and the emergence of rich spatio-temporal patterns.  We proposed a simple model build on the basis of paradigmatic models of active matter, synchronization and epidemic spreading. We performed a systematic exploration of the latter and provided a throughout characterization of its large-scale behavior, that we rationalized in several phase diagrams. As such, this study bridges together soft active matter physics and agent based modeling of complex systems,  pushing forward the knowledge boundaries of both fields by providing a reference framework to describe a plethora of natural and artificial systems, such as animal groups, crowds of people or robot swarms. 

Our model captures two local interaction mechanisms, based on the Kuramoto  and the SIS models. The former favors the emergence of synchronization, or flocking, while the latter the diffusion of information among the population.  Their interplay has a qualitative impact on the emergence of order: both flocking and information spreading can be achieved in regions of the phase space that would remain disordered in the absence of the feedback between these two.  
Strikingly, the SIS dynamics is able to drive a flocking phase transition even in the absence of explicit velocity-alignment interaction (i.e. $K = 0$), when the agents motion is completely determined by the epidemic process. In this scenario, the epidemic probability $\lambda$ controls a phase transition from a disordered state to an endemic polar one,  inducing the emergence of  collective motion. Such flocking transition induced by information spreading is found to be  discontinuous, characterized by the presence of hysteresis and metastability that we identified in the $\lambda - H$ plane, and followed by  the formation of traveling bands, recovering a behavior reminiscent of Vicsek-like systems, yet from a different perspective. 

The synchronized motion of the agents is generically accompanied by the emergence of swarms, i.e. dense structures moving coherently. In the $K = 0$ case, the agent's velocity $v_0$ crucially controls the threshold and the nature of the phase transition: as $v_0$ increases, the threshold $\lambda_c$ increases and the flocking transition changes from discontinuous to continuous. For large $v_0$, hysteresis, and the associated metastable region, disappears. The ordering mechanism in the fast and slow velocity regimes is thus qualitatively different, as  shown by the system's morphology: while at low velocities the system de-mixes into a high-density swarm and a low density incoherent background, at larger velocities the ordered phase consists on a homogenous flock. 
Interestingly, the nature of phase coexistence in our model is closer to the standard liquid-gas coexistence than the micro-phase separation characterizing the Vicsek model. The width and density of the traveling bands, induced by the SIS process, can be controlled by the velocity parameter $v_0$, which, consistently,   also controls the crossover from a discontinuous to a continuous phase transition behavior. 

The explanation of the role played by $v_0$ on the flocking transition induced by information spreading  is to be found, again, in the feedback mechanism between the agents motility and the SIS dynamics.  
The slower the agents move, the denser the swarms generated, fostering information diffusion across the swarm and therefore shifting the onset of flocking to lower values of the epidemic spreading probability. At low velocities, indeed, the interactions of neighboring agents favor the epidemic process and thus their coordinated motion. Two aligned agents moving together through space interact for longer time, fostering information spreading which, in turn,  promotes flocking.    At large velocities  such positive feedback  is broken, particles move faster than  information spreads, and the formation of local structures is impossible,  increasing the epidemic threshold. Previous epidemic models of mobile agents in which the agents mobility is decoupled from the epidemic dynamics, on the contrary, have reported larger epidemic threshold with respect to the mean field or homogeneous mixing case \citep{buscarino2008disease}, showing once again the crucial impact of the motility when coupled to information spreading. 



Our findings open several paths of future research. 
For instance, an analytic approach can be followed to study a simpler  limit case of our model, as shown in the Appendix for the dilute limit. The case in which the velocity-alignment is removed ($K=0$) and the feedback between mobility and the SIS process is broken ($H=0$) seems promising in this respect. Here we have shown that removing the feedback loop between mobility and SIS dynamics (i.e. decreasing $H$) leads to an increase of the epidemic threshold, but we did not specifically address the limit $H\to0$. In order to study analytically this case and quantify the increase of the epidemic threshold, the introduction of a noise term may be useful, since otherwise the agents motion would be  deterministic.

It is also worth noting that, in the present work, we  considered the motion of the agents and the SIS epidemic process as being synchronous (the integration of the dynamics in the real space and the space of the internal states is done without delay using the same time step). A time scale separation between the two dynamics may be considered, e.g. the SIS dynamics could be assumed to be much slower than the motion of the agents and the subsequently velocity-alignment. Such time scale factor cannot be explicitly absorbed in the definition of the other time scales such as $K$ or $H$, thus calling for future work. However, preliminary numerical explorations show no qualitative difference with the synchronous case in the long-time regime.

The static limit ($v_0 \to 0$), for which agents sit on a static random geometric graph, corresponds to a qualitatively different model that deserves a specific research effort. Very recent studies addressing Vicsek-like interactions on static networks show that the topology of the underlying network impacts the synchronization dynamics \cite{Romu}. In a static network, no dense structures that enhance epidemic spreading can be formed. However, the interplay between Kuramoto interaction and SIS dynamics can have effects on both synchronization and epidemic thresholds. 

Finally, in future work we aim at studying the response of our model to an external perturbation, in order to gain a better understanding of the spectacular collective reorganization events of animal groups facing a predator attack \cite{cavagna2010scale,procaccini2011,handegard2012dynamics}. 
Overall, our results shed light upon the effects of the interplay between information spreading and motility, and may constitute a guideline to design strategies for populations of artificial mobile agents (such as centimeter-sized robots \cite{Rubenstein2014,slavkov2018}) with targeted functionalities.

\section*{Acknowledgements}
A.D.-G. acknowledges MINECO for financial support under project FIS2015-71582-C2-2-P.
I. P. acknowledges MINECO and DURSI for financial support under projects FIS2015-67837- P and 2017SGR-884, respectively. 
M.S. acknowledges financial support by the J. McDonnell Foundation. 

\section*{Appendix A. Limiting cases of the model} 

Here we discuss in details several limiting regimes of the model, which allows gaining more  intuition on the dynamics of the system, and establishing connections with other models present in the literature.
We start by analyzing the motion of a single and free particle. 
In this regime, regardless of its initial epidemic state, the agent will spontaneously decay to its susceptible state (as long as $\mu>0$). Its dynamics is governed by 
 \begin{align}
&\boldsymbol{r}_i^{t+dt}=\boldsymbol{r}_i^t+v_0\,\boldsymbol{p}_i^t \, dt, \\ \nonumber
&\theta_i^{t+dt}=\theta_i^{t}+H \sin(\phi_i^t-\theta_i^t) \, dt \\ \nonumber
&\phi_i^{t+dt}=\phi_i^{t}+\delta \phi_i^t \, dt \\ \nonumber
 \end{align}
For a  non-zero value of $D_0$ and $H$, the agents follow a persistent random walk,  reminiscent of the paradigmatic Active Brownian Particles (ABP) model  but with a non-linear orientational noise term instead of the standard additive one. At short times,  $t\ll (D_0^{-1},\,H^{-1})$, the agent moves ballistically at  velocity $v_0$, while at longer time scales $t\gg (D_0^{-1}, \,H^{-1})$, its orientation becomes uncorrelated and its motion  diffusive. The parameter $D_0$ sets the reorientation rate of the phases, while $H$ introduces an extra time-scale, quantifying the typical time needed for an agent to adjust its orientation with its internal phase: $H$ plays the role of a damping coefficient. 

  \begin{figure}[tbp]
  \centering
    \includegraphics[scale=0.3,angle=0]{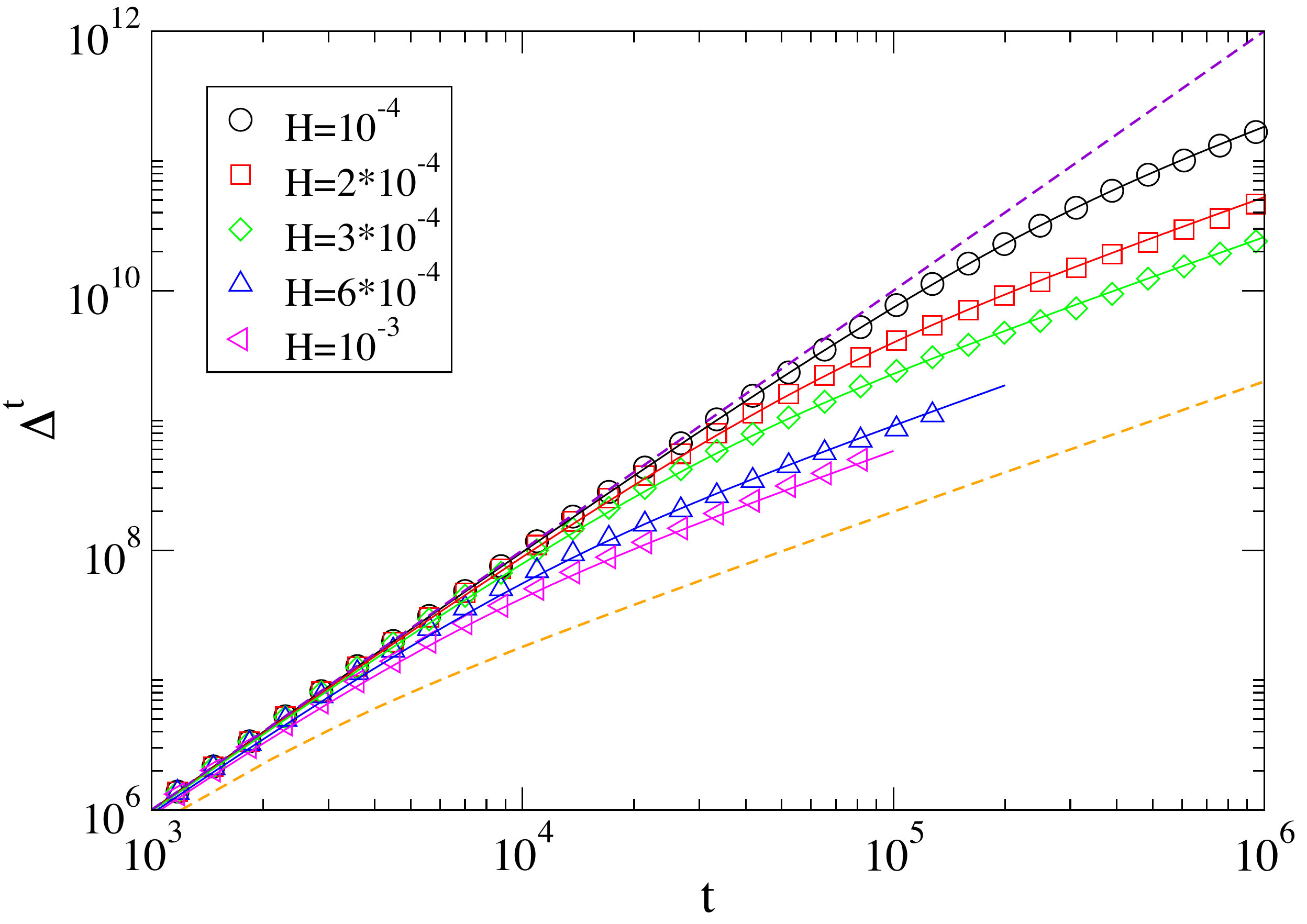}
    \includegraphics[scale=0.27,angle=0]{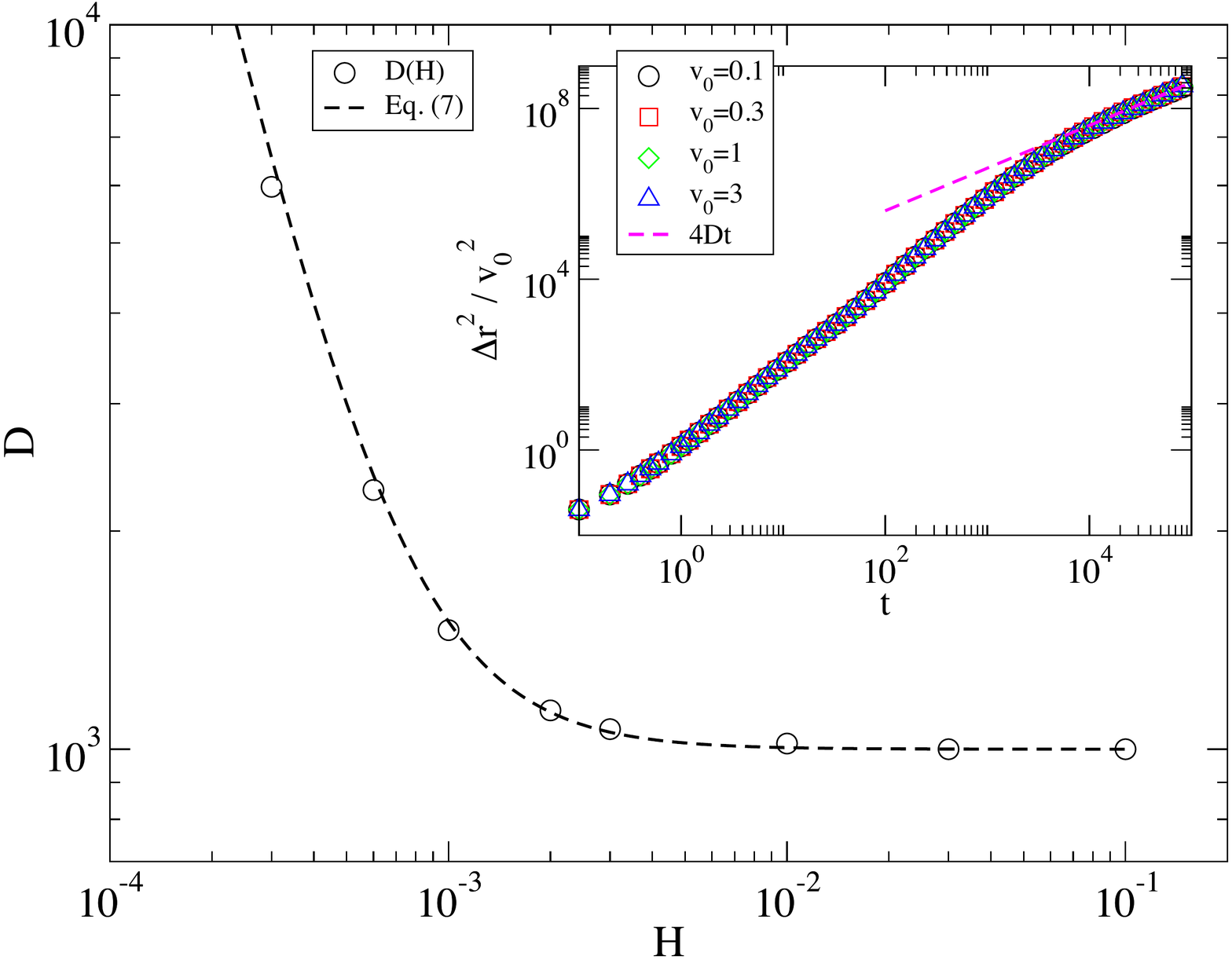}
          \caption{\textbf{Dilute limit.} Top: Mean-square displacement at fixed $v_0=1$ for different values of $H$ as given by eq. \ref{eq:msd} (represented in continuous lines) and obtained from simulations (symbols). Dotted lines correspond to the ballistic and ABP limit regimes, $H=0$ and $H\to\infty$, respectively.  Bottom: Diffusion coefficient $D$ for different values of $H$ at fixed $v_0=1$. The continuous line shows the value predicted by eqs. \ref{eq:msd}, \ref{eq:tau}.  The inset shows the mean-square displacement normalized by $v_0^2$ at fixed $H=0.1$ for different values of $v_0$. The linear growth $4Dt$, with $D$ given by  eq.  \ref{eq:tau} is also shown. }
       \label{fig:diff}
\end{figure}

Such behavior is illustrated in Fig. \ref{fig:diff} (a), showing the mean-squared displacement
$$ \Delta^t=\sum_i \frac{1}{ N}\langle[ \boldsymbol{r}_i^{t+t_0}-\boldsymbol{r}_i^{t_0}]^2\rangle$$
 for several values of $H$ at fixed $v_0$. 
At short times $\Delta^t=(v_0t)^2$ until the diffusive regime sets in and $\Delta^t=4Dt$ ($D$ being the diffusion coefficient). As shown in  Fig. \ref{fig:diff} (a), the mean-squared displacement of a single agent is well reproduced by the following expression 
 \begin{equation} 
  \Delta^t(t)= 2v_0^2\tau\left[t+\tau (e^{-t/\tau}-1) \right] \label{eq:msd}
  \end{equation}
 where $\tau$ quantifies the persistence time, i.e. the average time during which an agent moves ballistically, and it is given by 
 \begin{equation}
\tau(v_0,H)=\frac{1}{2D_0}+\frac{2D_0}{H^2} \,.\label{eq:tau}
\end{equation}
This is further confirmed by the direct measurement of  $D$,  defined by the long time behavior of the mean-squared displacement,  
as a function of $H$ (see Fig. \ref{fig:diff} (b)).  
In the limit of $H\to0$, the self-propulsion direction never adjusts to the internal phase, and the agent follows a rectilinear and uniform motion yielding $\tau\to\infty$. In the $H\to\infty$ limit the direction of self-propulsion is synchronous to its internal phase, therefore,  agents are free Active Brownian Particles whose diffusion is solely controlled by the rotational noise of strength $D_0$ and $D=v_0^2/(2D_0)$. 
In general, it is the combination of both time scales, $1/D_0$ and $1/H$ which determines the motion of a free agent. 
 

In the limit $H\to0$, the SIS process is still affected by the agents' mobility, but it no longer affects the agents' mobility. 
The self-propulsion direction $\theta$ is completely determined by the Kuramoto-like interaction (see eq. \eqref{eq:theta_ev}), irrespective of the internal phase  $\phi$. The feedback between the internal epidemic state and the motion of the agent is lost. 
As opposed to models of flocking which explore the competition between aligning interactions and noise,  our model is deterministic in this limit,
since the noise is introduced through the dynamics of the internal phases $\phi$. 

In the limit $\lambda\to0$, the epidemic process stops, so the internal phases $\{\phi\}$ follow free diffusion. Therefore, the variables $\{\theta\}$ follow a (noisy) Kuramoto dynamics with local coupling. In turn, the agents self-propel and align their velocities with their neighbors; like in the Vicsek model  but where noise is introduced via a sine, see Eq. \eqref{eq:theta_ev}. 
In this limit, we expect to reproduce the main phenomenology of Vicsek-like models: below some threshold $K_c$ we expect the system to be disordered, while for $K>K_c$ we expect to observe flocking, the collective motion of a macroscopic fraction of the system, or, equivalently, the global synchronization of the agents' velocities.

\bibliography{sisbis}{}

\end{document}